\documentclass[reprint,bibnotes,amsmath,amssymb,showpacs,floatfix,superscriptaddress,bibliography]{revtex4-2}

\usepackage{amsmath,empheq}
\usepackage{amsfonts}
\usepackage{amssymb}
\usepackage{amsxtra}
\usepackage{mathtools}
\usepackage{xcolor}
\usepackage{graphicx}
\usepackage{subfigure}
\usepackage{dcolumn}
\usepackage{mathrsfs}
\usepackage{tikz}
\usepackage{float}
\usepackage{bm}
\usepackage[breaklinks=true,colorlinks,citecolor=blue,linkcolor=blue,urlcolor=blue]{hyperref}

\DeclareMathAlphabet{\bi}{OML}{cmm}{b}{it}
\def\be{\begin{equation}}
	\def\ee{\end{equation}}
\def\bearr{\begin{eqnarray}}
	\def\eearr{\end{eqnarray}}
\def\la{\langle}
\def\ra{\rangle}

\begin{document}
  	\title{Valley resolved optical spectroscopy and coherent excitation of quantum Hall edge states in graphene}
	\author{Ashutosh Singh}
	\email{asingh.n19@gmail.com}
	\affiliation{Department of Physics and Astronomy, Texas A\&M University, College Station, TX, 77843 USA}
 \author{Maria Sebastian}
\affiliation{Department of Physics and Astronomy, Texas A\&M University, College Station, TX, 77843 USA}
\author{Mikhail Tokman}
\affiliation{Department of Electrical and Electronic Engineering and Schlesinger Knowledge Center for Compact Accelerators
and Radiation Sources, Ariel University, 40700 Ariel, Israel}

\author{Alexey Belyanin}%
\email{belyanin@tamu.edu}
\affiliation{Department of Physics and Astronomy, Texas A\&M University, College Station, TX, 77843 USA}

	\date{\today}

    \begin{abstract}
		We show that chiral edge states in graphene under Quantum Hall effect conditions can be selectively probed and excited by terahertz or infrared radiation with single-quasiparticle sensitivity without affecting bulk states. Moreover, valley-selective excitation of edge states is possible with high fidelity. The underlying physical mechanism is the inevitable violation of adiabaticity and inversion symmetry breaking for electron states near the edge. This leads to the formation of Landau level-specific and valley-specific absorbance spectral peaks that are spectrally well separated from each other and from absorption by the bulk states, and have different polarization selection rules. Furthermore, inversion symmetry breaking enables coherent driving of chiral edge photocurrents due to second-order nonlinear optical rectification which becomes allowed in the electric dipole approximation.

        
	\end{abstract}
    
\maketitle
\section{Introduction}

 Graphene provides a superior and versatile platform for the Quantum Hall (QH) effect studies \cite{Kim}, due to the high mobility of both types of carriers \cite{BOLOTIN2008351}, an easier fabrication, gating, and control of the Fermi level, possibility of lateral p-n junctions \cite{Levitov} and valley-selective transport \cite{Xiao,Komatsu}, as well as new opportunities arising from the formation of bilayers and hybrid van-der Waals structures\cite{Yankowitz, Andrea}. 

As with any topological system, a major focus in the QH research has been the interplay between the electron states in the insulating bulk of the sample and the current-carrying chiral edge states  \cite{Halperin_Edge, MacDonald_edge} which have been extensively studied with real-space imaging and momentum-resolved electron spectroscopy \cite{PhysRevX.4.011014,Marguerite2019, PhysRevB.107.115426,Li2013, Patlatiuk2018, Kim2021}. Chiral edge states in graphene have a number of unique features such as the presence of two types of edges with distinct properties, zigzag and armchair \cite{PhysRevB.73.195408}. The coherence of unidirectional electron transport in QH edge states has stimulated a massive amount of  research in QH edge state interferometry. Various types of electron interferometers have been implemented, both in conventional semiconductor quantum wells and in graphene, and for both integer and fractional statistics of carriers \cite{PhysRevLett.62.2523, Chamon, Ji2003, PhysRevB.73.245311, PhysRevLett.96.016802,PhysRevLett.97.186803, PhysRevB.79.241304,PhysRevB.82.085321,Corentin2021, PhysRevB.105.165310, Nakamura2020, Ron2021, nak23}. 

\begin{figure}[ht!]
\includegraphics[width =.9\linewidth]{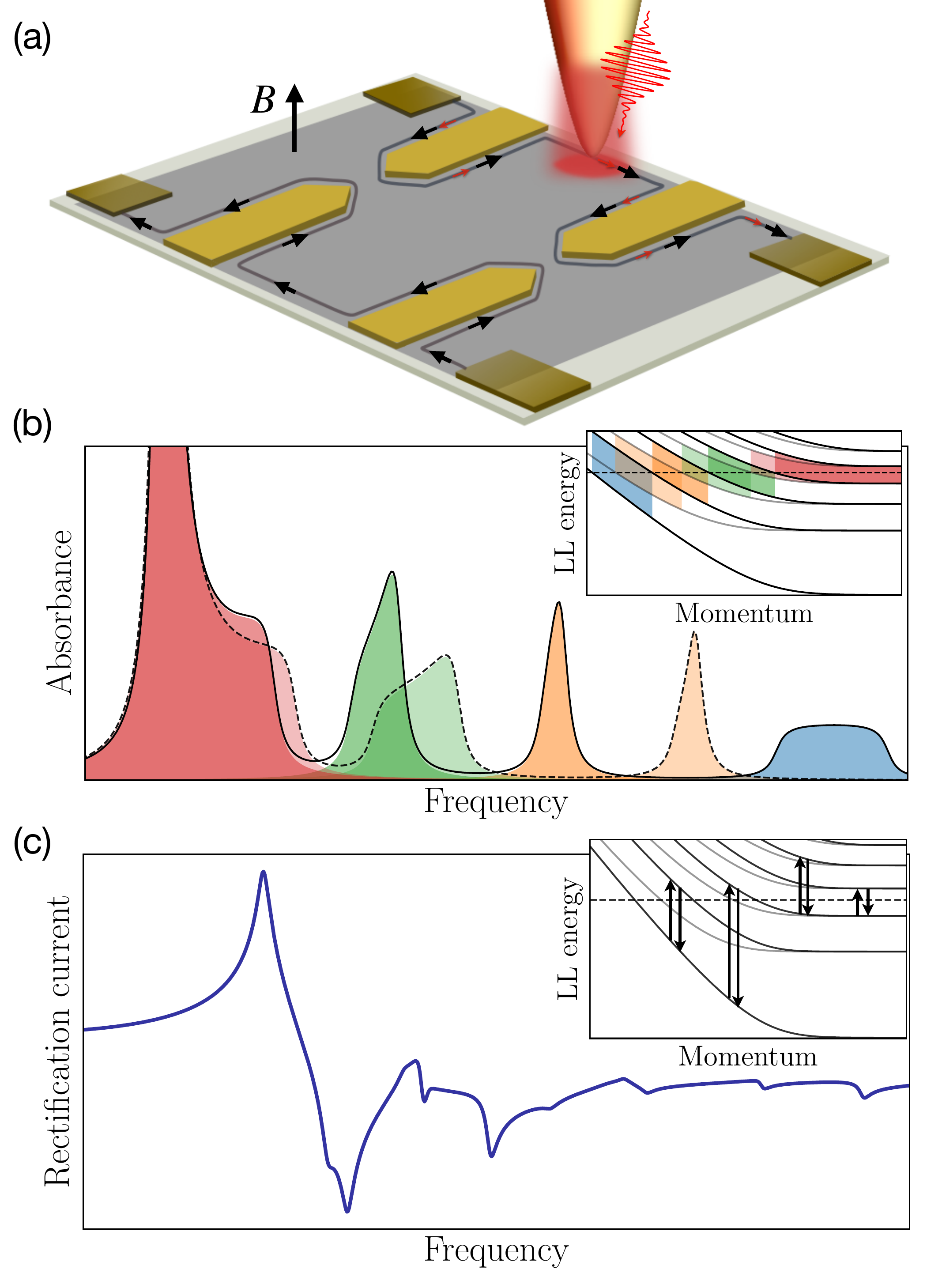}
\caption{(a) Schematic of a nano-tip enabled optical excitation of chiral edge states  in a typical quantum Hall interferometer. (b) An example of 2D absorbance spectra showing well separated spectral peaks corresponding to the excitation of specific edge states and different valleys: K-valley states (solid line) and K'-valley (dashed line). Inset: dispersion  of electron eigenenergies and optical transitions corresponding to different part of the absorption spectra, indicated by the same color coding.  Dashed line denotes the chemical potential below which all states are occupied. (c) An example of DC photocurrent spectrum resulting from the second-order nonlinear optical rectification process in the electric dipole approximation. Inset: a segment of electron eigenenergies, with a few examples of the optical transitions contributing to the optical rectification. The current is obtained by integrating over electron momenta and summing up over all states for a given Fermi level (dashed line).   
}
\label{fig1}
\end{figure}
In contrast to purely transport experiments, optical spectroscopy and pumping of QH states in graphene provides an interesting opportunity of probing excited electron states or generating photocurrents \cite{PhysRevB.53.1054, Hafezi}. 
Far-infrared radiation permits resonant excitation or spectroscopy  of low-lying Landau-quantized electron states in graphene \cite{PhysRevLett.97.266405, PhysRevLett.98.197403}, including resonantly enhanced nonlinear response \cite{PhysRevLett.108.255503, PhysRevLett.110.077404}. If the goal is probing electron states under QH effect conditions, one should still exercise caution. Resonant optical transitions between bulk Landau levels can create  nonequilibrium carrier population and nonzero bulk DC conductivity across the sample. In fact, the modification of the magnetotransport in the QH samples by sufficiently strong resonant microwave or terahertz fields is a well-developed area of research; see, for example, the review in \cite{RevModPhys.84.1709,Kriisa2019} and some of the recent papers \cite{PhysRevB.96.115449,PhysRevB.106.L161408}. 

In the recent paper \cite{singh}, we demonstrated that optical spectroscopy and coherent optical control of specific QH edge states for nonrelativistic electrons in semiconductor quantum wells (QWs) is possible and can be in fact highly effective without disrupting the QH effect conditions in the bulk. The key physical reason behind this lies in the fact that the optical transitions between electron states near the sample boundary (within a few magnetic lengths from the edge) have significantly higher transition energies and different selection rules compared to the bulk of the sample. This enables excitations of edge states with single quasiparticle sensitivity without affecting the bulk states. 

In this paper we focus on the optical excitation of QH chiral edge states for massless Dirac electrons in graphene. The graphene edge states offer a greater variety of optical transitions due to two kinds of edge terminations. More importantly, their optical response has a number of unique features. Perhaps the most interesting of them is the possibility of valley-selective excitation. As shown in Fig.~\ref{fig1} and also discussed in greater detail below, the absorption peaks corresponding to excitation of edge states near K and K' valleys are spectrally well separated from each other and from the absorption by bulk states.  Moreover, by utilizing different polarization selection rules for edge states and subtracting between two orthogonal polarizations, one can eliminate any residual bulk state contribution. 

Furthermore, inversion symmetry breaking near the sample boundary enables strong second-order optical nonlinearity in the electric dipole approximation, resulting in efficient optical rectification of incident radiation and direct optical driving of a quasi-DC current in specific edge states.  The spectrum of the optically driven DC current and the examples of the optical transitions contributing to the rectification are sketched in Fig.~\ref{fig1}(c), with quantitative discussion provided in the subsequent sections. 

While resonant edge-state excitation has excellent energy and momentum selectivity, it suffers from size mismatch generic to the THz spectroscopy: namely, the size of the QH sample can be as small as several micrometers whereas the wavelength scale is in hundreds of micrometers. If the goal is to excite or probe a particular edge of the sample or a given arm of the interferometer. one should use an illuminated nanotip as sketched in Fig.~\ref{fig1}(a) or integrate the sample with a nanoantenna or nanocavity. This has been done in many THz experiments; e.g., \cite{10.1063/1.2938416, 10.1063/1.1476061, 10.1063/5.0189061}, although here we do not consider the possibility of strong or ultrastrong coupling of Landau level polaritons in a cavity, which has recently become yet another popular research area \cite{doi:10.1126/science.abl5818, PhysRevB.111.L121407}.  

The high spatiotemporal and energy selectivity of optical excitations of chiral edge states provides a sensitive spectroscopy tool complementary to electron transport measurements and enables coherent control of individual edge channels in QH interferometers, endowing them with new optoelectronic functionality. Therefore, we hope that our paper will stimulate further collaboration between optical and QH effect communities.

The structure of the paper is as follows: In Section \ref{CES}, we provide the eigenstates and eigenenergies for electrons near the zigzag edge. 
In Section \ref{Absorp_prob} we calculate the single-photon absorption probability for a quantized optical field and derive the 2D absorbance spectra which reveal sharp characteristic peaks at high frequencies that are entirely due to nonadiabatic edge states near zigzag edge. Unlike the case of nonrelativistic electrons \cite{singh}, here in graphene the spectral peaks corresponding to specific edge states and specific valleys are well-isolated from each other. Therefore, valley-selective excitation of specific edge states is possible with high fidelity.

Section \ref{DC_gen} calculates the spectra of the DC current originated from the second-order nonlinear process of the optical rectification in the electric dipole approximation.  In Sec.~\ref{armchair} we derive the eigenstates, transition matrix elements, and 2D absorbance spectra in the case of the armchair edge. We do not show the results for the nonlinear current in the armchair edge case because it looks qualitatively similar to the zigzag edge. The Conclusions summarize the results and some of their implications.  

\section{Quantum Hall edge states in graphene}\label{CES}
\subsection{Energy dispersion}
The charge dynamics for low energy quasiparticles in graphene can be described by the Hamiltonian  
\begin{align}\label{gr_dispersion}
H_{\bf k} = \hbar v_F{\boldsymbol\sigma}\cdot{\bf k}~,
\end{align}
where $v_F \approx 10^8$ cm/s is the Fermi velocity, ${\boldsymbol\sigma}$ is the Pauli matrix triplet and ${\bf k} = (k_x, k_y)$ is the two-dimensional quasiparticle momentum counted from the K-point. The Hamiltonian near the K'-point is obtained by $k_x \rightarrow -k_x$. 
In the presence of a quantizing magnetic field $B$ along $z$-axis, described by the vector potential ${\vec {\mathcal A}} = -y B \hat x$, the Hamiltonian becomes
\begin{align}
H_{\bf k} = \hbar v_F\left(\sigma_x k_x + \sigma_y k_y\right) - \frac{eBy}{c} v_F\sigma_x~,
\end{align}
giving rise to a discrete set of eigenstates and eigenenergies for a given $k_x$, hereafter denoted simply as $k$. The corresponding eigenvalue problem must be solved with proper boundary conditions. 
We define a spinor $\Psi(x,y) \sim e^{i kx}\left(\Phi^{(1)}(y), \Phi^{(2)}(y)\right)^{\rm T}$, such that the
Schr\"{o}dinger equation for the $K$-valley becomes 
\begin{align}
\frac{\partial^2\Phi^{(2)}}{\partial \tilde y^2} + \left(2E^2 + 1 - (\tilde y - \tilde y_k)^2\right)\Phi^{(2)} = 0\label{Phi_2}~,\\
\frac{\partial^2\Phi^{(1)}}{\partial \tilde y^2} + \left(2E^2 - 1 - (\tilde y - \tilde y_k)^2\right)\Phi^{(1)} = 0\label{Phi_1}~.
\end{align}
Here $ \Phi^{(1)}(y)$ and $\Phi^{(2)}(y)$ refer to the wave functions at sublattice $A$ and $B$ respectively, $E$ is the LL energy, $\tilde y = y/\ell_c$, the center of the cyclotron oscillator $\tilde y_k = y_k/\ell_c$ and $\ell_c = \sqrt{hc/(eB)}$ is the magnetic length. Unlike the case of a nonrelativistic two-dimensional electron gas (2DEG) in semiconductor QWs, here the treatment is different for different kinds of edge termination \cite{PhysRevB.73.195408, Gusynin2008, PhysRevB.83.045421, PhysRevResearch.3.013201}. For the zigzag edge, only one of the sublattice wave functions vanishes whereas for armchair edge both components vanish simultaneously. We start from the zigzag edge. To be specific, following \cite{Gusynin2008}, we assume that the physical sample terminates with atoms of sublattice $B$, so that at $y = 0$ the atoms of sublattice $A$ are already missing and we force the wavefunction at sublattice $A$ to go to zero there. The solution for $\Phi^{(1,2)}$ can be given in terms of the functions of the parabolic cylinder \cite{Gusynin2008,PhysRevB.83.045421, PhysRevResearch.3.013201}, such that the wavefunction for the $K$-valley is
\begin{align}\label{psi_K}
\Psi_{nk, s}^K(x,y) = \frac{e^{i kx}}{\sqrt{C_{n k}}}\begin{pmatrix}
        D_{n}(\sqrt{2}(y/\ell_c - k\ell_c))\\
        s\sqrt{n}D_{n-1}(\sqrt{2}(y/\ell_c - k\ell_c))
	\end{pmatrix},
\end{align}
where $C_{nk}$ are normalization constants, $s = +(-)$ refers to the conduction (valence) band and $k$ is measured from the $K$ point. Here $n$ denotes the LL index for the states far from the boundary, which becomes dependent on  momentum $k$ near the edge. Note that it makes the index of parabolic cylinder functions to be a non-integer, continuous function of $k$. The parabolic cylinder functions can be expressed in terms of Hermite polynomials in the bulk. The eigenenergies can be written as $E^s_{nk} = s\hbar\omega_c\sqrt{n}$, where $\omega_c = \sqrt{2} v_F/\ell_c$. For a magnetic field strength of $1$T, the magnetic length $\ell_c\approx 26$ nm, and the cyclotron frequency $\omega_c\approx 5.5\times 10^{13}$ s$^{-1}$ corresponding to the wavelength of 34 $\mu$m in the far-infrared.

For the $K^{\prime}$-valley, the wavefunctions can be obtained by operating $\sigma_z\sigma_x$ on the spinor in Eq.~(\ref{psi_K}) which changes the sign of $k_x$ in Eq.~\eqref{gr_dispersion} such that
\begin{align}\label{psi_Kp}
\Psi_{nk, s}^{K^{\prime}}(x,y) = \frac{e^{i kx}}{\sqrt{C_{n k}}}\begin{pmatrix}
        s \sqrt{n}D_{n-1}(\sqrt{2}(y/\ell_c - k\ell_c))\\
        -D_{n}(\sqrt{2}(y/\ell_c - k\ell_c))
	\end{pmatrix}.
\end{align}

In order to determine how $n$ changes with $k$, we use the fact that the electron wavefunction at one of the sublattices needs to vanish as $y\to 0$.
We set $D_{n}(\sqrt{2}(- k\ell_c)) = 0$ for the $K$-valley and $\sqrt{n}D_{n-1}(\sqrt{2}( - k\ell_c)) = 0$ for the $K^{\prime}$-valley respectively.
\begin{figure}[t!]
\includegraphics[width =\linewidth]{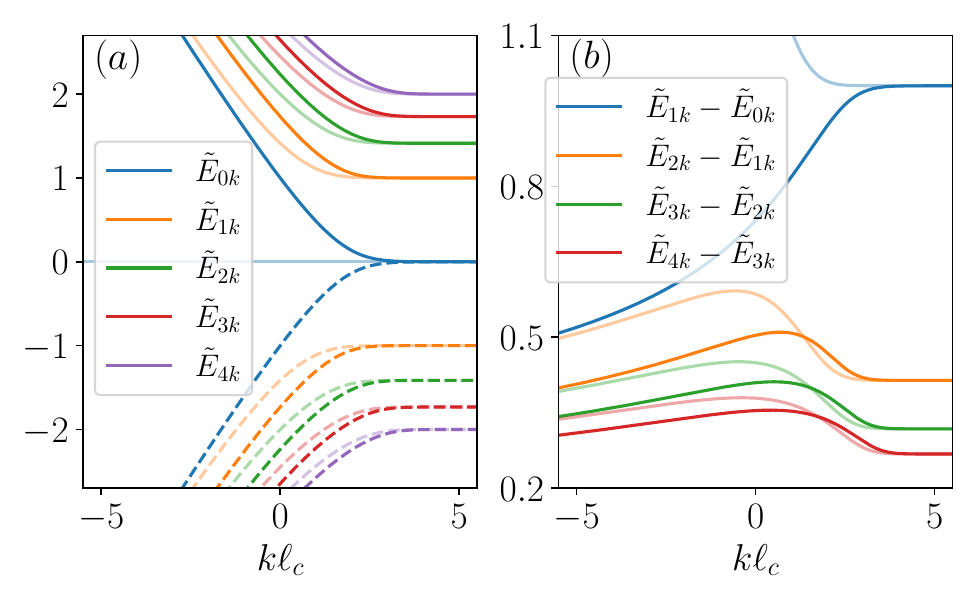}
\caption{(a) LL energies normalized by $\hbar\omega_c$ ($\tilde E_{nk} = E_{nk}/\hbar\omega_c$) for positive (solid) and negative (dashed) branches are shown as a function of $k\ell_c$ for $K$ (solid color) and $K^{\prime}$ (light color) valleys.  (b) Normalized optical transition energies between consecutive LLs as a function of $k\ell_c$ for the two valleys. For $k\ell_c\gg 0$, the values of the transition energies are $\tilde E_{1k}-\tilde E_{0k} = 1$, $\tilde E_{2k}-\tilde E_{1k} \approx 0.414$, $\tilde E_{3k}-\tilde E_{2k} \approx 0.32$, $\tilde E_{4k}-\tilde E_{3k} \approx 0.27$, and $\tilde E_{5k}-\tilde E_{4k} \approx 0.24$. }
\label{dispersion}
\end{figure}

The energy dispersion for the first five LLs in $K$ and $K^{\prime}$ valleys is shown in Fig.~\ref{dispersion}(a). For $k\ell_c\gg 0$, these LLs are dispersionless with their energies given as $0, \hbar\omega_c, \sqrt{2}\hbar\omega_c, \sqrt{3}\hbar\omega_c,...$. For states close to the edge, the energies become dispersive and at the same time the degeneracy between the $K$ and $K^{\prime}$ valleys is also lifted. Only the zeroth LL in $K^{\prime}$ valley remains dispersionless. This is due to the fact that $D_{n-1}(\sqrt{2}( - k\ell_c))$ remains finite for small $n$, so that $\sqrt{n}D_{n-1}(\sqrt{2}( - k\ell_c)) = 0$ implies $n = 0$.
Therefore in the $K^{\prime}$ valley, 
\begin{align}\label{psi_Kp_0}
\Psi_{0}^{K^{\prime}}(x,y) = \frac{e^{i kx}}{\sqrt{C_{0}}}\begin{pmatrix}
        0\\
        D_{0}(\sqrt{2}(y/\ell_c - k\ell_c))
	\end{pmatrix}.
\end{align}
No such dispersionless solution exists in the $K$ valley. 
%
%
In Fig.~\ref{dispersion}(b) we plot the transition energies between neighboring LLs as a function of electron momentum. While for nonrelativistic massive electron systems the LLs in the bulk are equidistant and the transition energies increase monotonically when approaching the edge \cite{singh},  the transition energies of electrons in graphene show non-monotonic behavior. This causes the photon absorption spectra to develop distinct peaks as we will see below. Furthermore, the valley degeneracy of the bulk states is lifted when approaching the edge. In contrast to the $K$-valley, the zeroth energy state in the $K^{\prime}$ valley remains dispersionless.
%

\subsection{Transition matrix elements between edge states}
\begin{figure}[ht!]
\includegraphics[width =\linewidth]{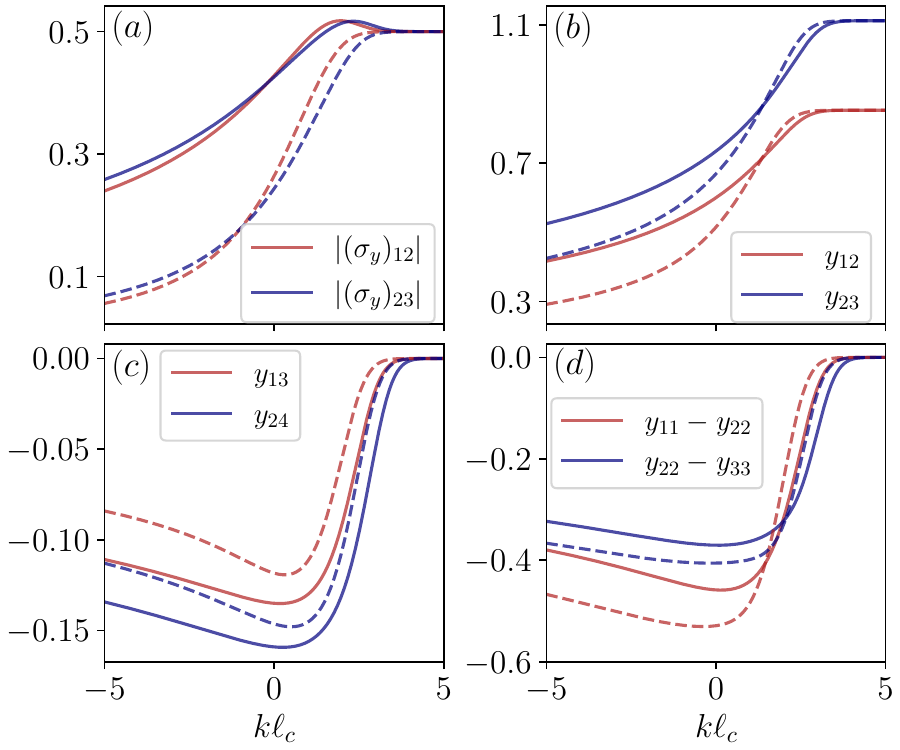}
\caption{Transition matrix elements involving states with positive energies for $K$ (solid) and $K^{\prime}$ (dashed) valley for $\hat y$-polarized electric field as a function of $k\ell_c$ . (a) $y$ component of velocity matrix elements (normalized by $v_F$)
and (b) $y$ component of the dipole matrix elements (normalized by $e\ell_c$)for optical transitions $1\rightarrow 2$ and $2\rightarrow 3$. 
(c) $y$ component of the dipole matrix elements for optical transitions $1\rightarrow 3$ and $2\rightarrow 4$ that are electric-dipole forbidden in the bulk. (d) Differences between permanent dipole matrix elements that are zero in the bulk.}
\label{dipole}
\end{figure}
Here we calculate the matrix elements of the velocity and position operators which determine the strength and selection rules of the optical transitions between electron states. For bulk states in graphene (far from the edges), the electric-dipole allowed transitions satisfy the selection rule $\Delta |n| = \pm 1$ for left and right circularly polarized light \cite{Falko, Yao_2013}. For the states near the edge where inversion symmetry is broken this is no longer the case. The 2D velocity operator is just the Pauli matrix vector, $v_F{\boldsymbol\sigma}$, and the general relationship  between velocity and position matrix elements is given by
\begin{equation}
{\bf r}_{\alpha \beta} = \frac{i \hbar {\bf v}_{\alpha \beta}}{E_{\alpha} - E_{\beta}},    
\end{equation}
or, after normalization, 
%
%
\begin{align}\label{r_sigma}
(\tilde {\bf r})^{ss^{\prime}}_{nmk} = \frac{i}{\sqrt{2}}\frac{({\boldsymbol\sigma})^{ss^{\prime}}_{nmk}}{\tilde E^s_{nk} - \tilde E^{s^{\prime}}_{mk}}~,
\end{align}
where, $(\tilde {\bf r})^{ss^{\prime}}_{nmk} = \langle \Psi_{nk, s}|{\bf r}|\Psi_{mk, s^{\prime}}\rangle/\ell_c$ and $({\boldsymbol\sigma})^{ss^{\prime}}_{nmk} = \langle \Psi_{nk, s}|{\boldsymbol\sigma}|\Psi_{mk, s^{\prime}}\rangle$.
For the chosen gauge, it is convenient to calculate the $y$-component (dimensionless) of the dipole matrix element,
\begin{align}\nonumber\label{dipole_mat}
\langle \Psi_{nk, s}|\tilde y|\Psi_{mk, s^{\prime}}\rangle = \int_{0}^{\tilde y_{\rm max}} \tilde yd\tilde y  \left[D_n(y^{\prime})D_m(y^{\prime}) \right.\\
\left. + ss^{\prime}\sqrt{nm}D_{n-1}(y^{\prime})D_{m-1}(y^{\prime}) \right]~,
\end{align}
where we have defined $y^{\prime} = \sqrt{2}(y/\ell_c - k\ell_c)$. 
For the relevant range of momenta the wave functions vanish quickly with increasing $y$ so that $\tilde{y}_{\rm max} = 20$ is large enough. 
As a consistency check for our numerical solution, we compared with the $y$-component of Eq.~\eqref{r_sigma} which match well for both bulk and edge states. 

Fig.~\ref{dipole} shows several examples of the $y$-components of the normalized velocity (a), position (b), (c) and difference between position (d) matrix elements. Parts (a) and (b) of the figure show the transitions electric dipole-allowed in the bulk. With approaching the edge, the transition element magnitudes are split between the valleys. The normalized velocity matrix elements saturate to $0.5$ for $k\ell_c \gg 0$ and decrease monotonically for states with increasingly negative momentum for the $K^{\prime}$ valley whereas for the $K$ valley these values increase initially before decreasing as momentum changes from the positive to negative values. Parts (c) and (d) show the components that are forbidden in the bulk but become electric-dipole allowed due to inversion symmetry breaking near the edge. They represent several of the many 
terms contributing to the second order nonlinear response due to edge states, which is electric-dipole forbidden in the bulk. 

%
\begin{figure}[t!]
\includegraphics[width =\linewidth]{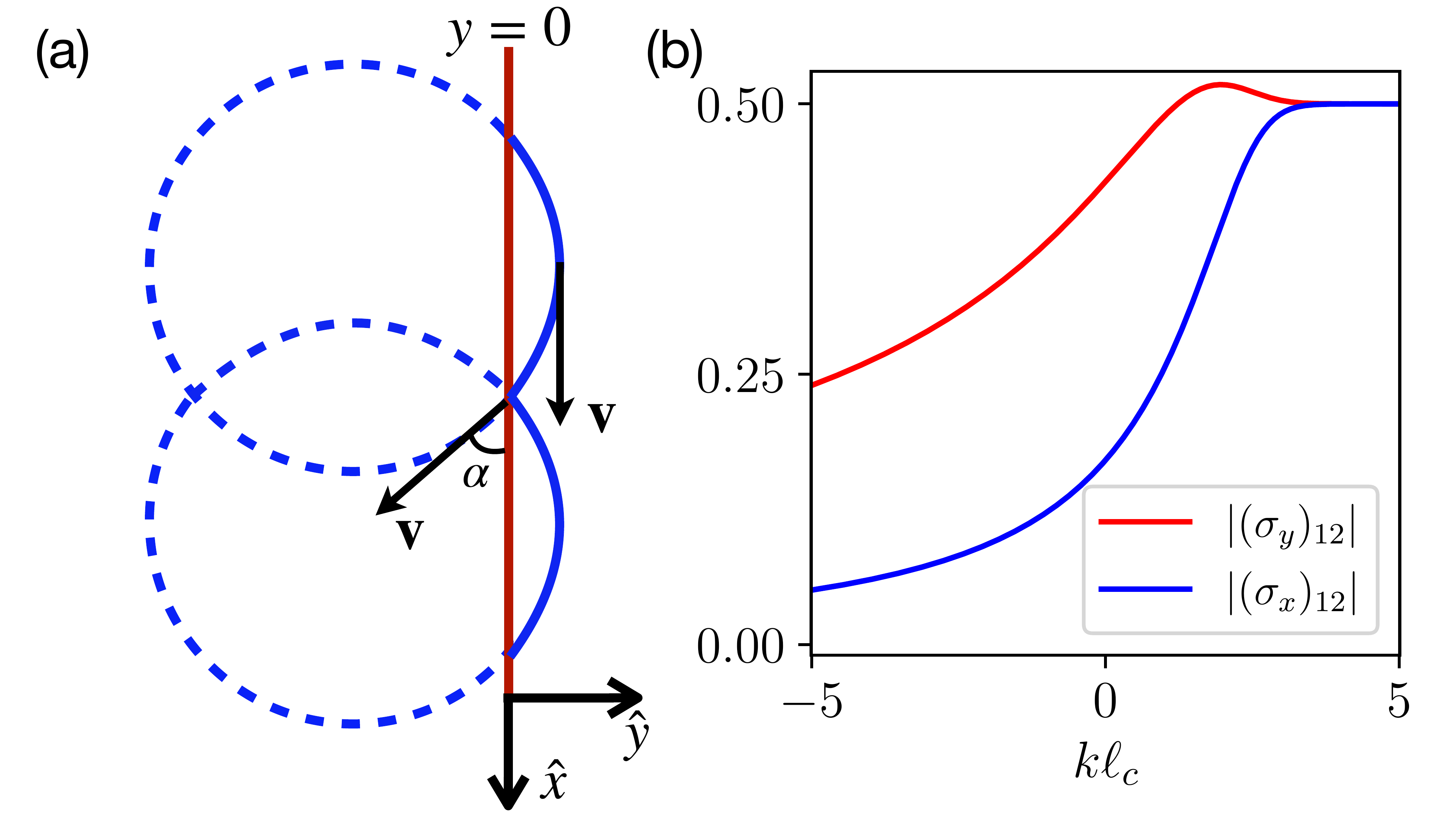}
\caption{(a)  Classical picture for electron motion near the edge and (b) $\hat x$ and $\hat y$ components of the matrix elements  of the normalized velocity ${\boldsymbol\sigma}$ for the LL $n = 1$ to $n = 2$ transition.}
\label{dipole_pol}
\end{figure}

Figure \ref{dipole_pol} illustrates another peculiarity of the optical transitions resulting from inversion symmetry breaking and squeezing of electron wave functions near the edge: the modification of the polarization selection rules. Namely, the $x$-component of the transition matrix elements becomes suppressed  and the transitions become increasingly elliptically polarized; see part (b) of the figure. This is a universal feature which exists also for massive nonrelativistic electrons \cite{singh}, and in fact has a quasiclassical origin. Indeed, as illustrated in part (a), when approaching the edge $y = 0$ (i.e., at large negative $y_k = k \ell_c^2 < 0$) the classical trajectory becomes increasingly aligned along $x$ and squeezed in $y$-direction, but the change in velocity is greater along $y$: $\Delta v_y = v \sin\alpha \sim v \alpha \gg \Delta v_x = v(1-\cos\alpha) \sim v \alpha^2/2$.  

%
%
%
\section{Absorbance spectra}\label{Absorp_prob}


Peculiar, valley-selective electron dispersion near the edge, shown in Fig.~\ref{dispersion}, in combination with modified transition matrix elements and polarization selection rules promises unique optical response and the possibility to resonantly excite a single electron into a given edge state and in a given valley. Indeed, here we show this to be the case. We calculate the single-photon absorption probability, which also gives dimensionless 2D absorbance spectra. Assuming the Fermi level in the conduction band, we will include only the transitions between the conduction band states. Interband transitions are at higher, mid-infrared energies which could be an advantage in the experiment, but they have significantly lower dipole matrix elements and weaker absorption. 

We need to keep the electron-photon coupling description fully quantized. 
The quantum state of electrons can be represented in terms of occupation numbers of $|nk\rangle$ states, for example $| \cdots 1_{nk} \cdots 0_{n'k'} \cdots \rangle $, where $1_{nk}$ and $0_{n'k'}$  are occupied and unoccupied states. Fermionic annihilation and creation operators are acting on these states in the usual way: $\hat{a}_{nk} | \cdots 1_{nk} \cdots \rangle = | \cdots 0_{nk} \cdots \rangle$ and $\hat{a}_{nk}^{\dagger} | \cdots 0_{nk} \cdots \rangle = | \cdots 1_{nk} \cdots \rangle$.  The electron Hamiltonian is then $\hat{H}_e = \sum_{n,k}E_{nk}\hat{a}_{nk}^{\dagger}\hat{a}_{nk}$.

%
\begin{figure*}[ht!]
\includegraphics[width =\linewidth]{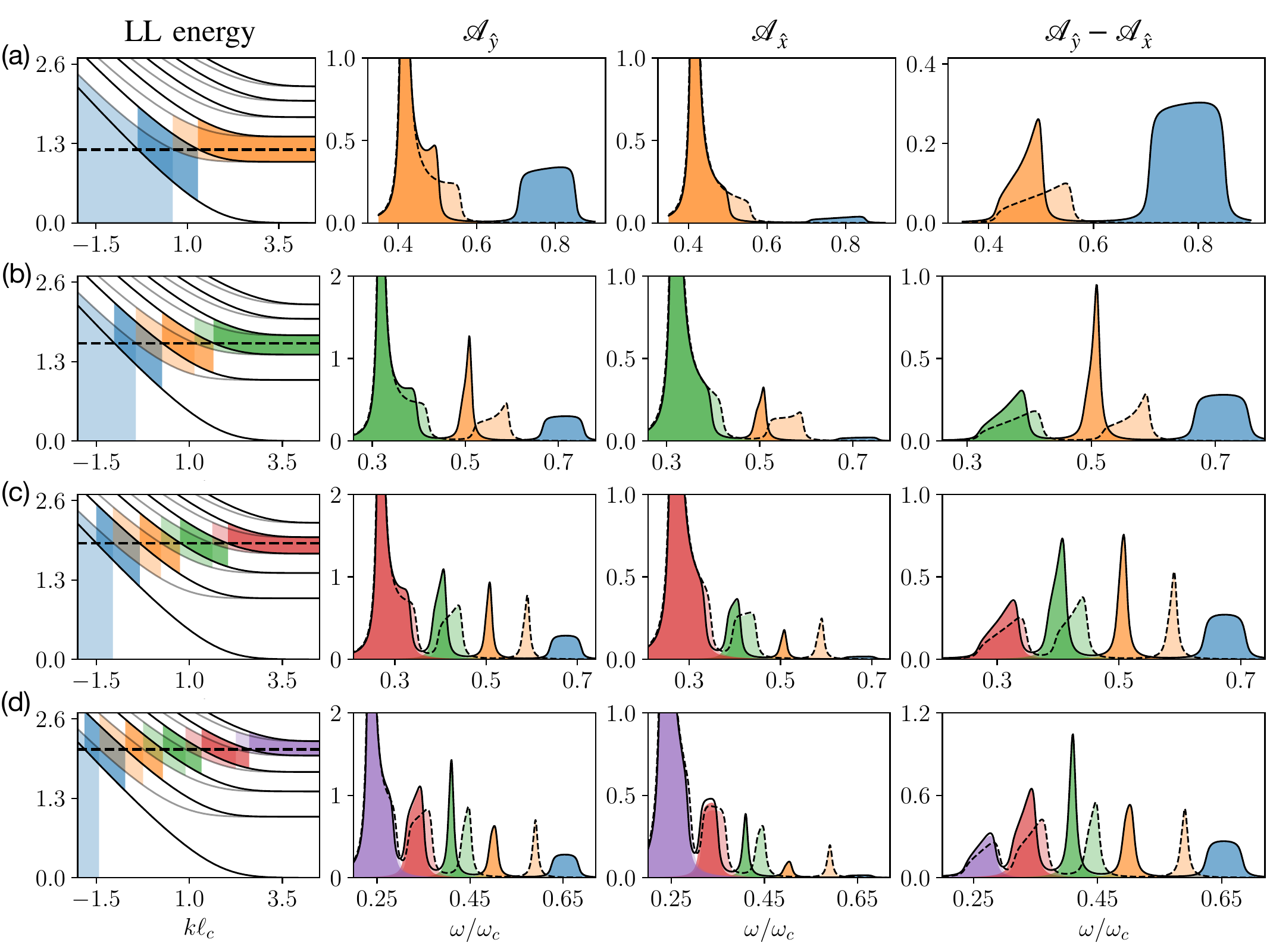}
\caption{Geometry independent absorbance ${\mathscr A}_{{\bf e}_{q}}(\omega)$ for $\hat y$ and $\hat x$ polarized electric field and their differences for $K$ (solid black) and $K^{\prime}$ (dashed black) valley as a function of normalized optical frequency $\omega/\omega_c$ for different Fermi levels: (a) $E_F = 1.2\hbar\omega_c$, (b) $E_F = 1.6\hbar\omega_c$, (c) $E_F = 1.9\hbar\omega_c$ and (d) $E_F = 2.1\hbar\omega_c$. The decay rate is taken to be $\gamma = 0.005\omega_c$. First column: electron energy dispersion near the edge and the Fermi level (dashed line). The regions contributing to different absorbance peaks are shaded with colors that match the colors of the peaks. Second column: absorbance of $y$-polarized light;  third column: absorbance of $x$-polarized light; fourth column: difference between absorbances of $y$ and $x$-polarized radiation. Solid and dashed curves indicate the absorbance due to $K$ and $K^{\prime}$-valley states, respectively. 
}
\label{absorbance_KKp}
\end{figure*}
%

The Hamiltonian and the operator of the vector potential of the EM field incident on the sample are 
\begin{equation}
\hat{H}_{\rm ph} = \sum_{n,{\bf q}}\hbar\omega_{n {\bf q}}\left(\hat{b}_{n {\bf q}}^{\dagger}\hat{b}_{n {\bf q}} + \frac{1}{2}\right),
\end{equation}
\begin{align}\label{A_field}
\hat {\bf A} = \sum_{n, \mathbf{q} }\sqrt{\frac{2\pi c^2\hbar}{V\omega_{\bf q}}}\left({\bf e}_{n{\bf q}}\hat{b}_{n {\bf q}}e^{i{\bf q}{\bf r}} + {\bf e}^{*}_{n{\bf q}}\hat{b}^{\dagger}_{n {\bf q}}e^{-i{\bf q}{\bf r}}\right)~,
\end{align}
where $\hat{b}_{n \mathbf{q} }$ and $ \hat{b}^{\dagger}_{n\mathbf{q} }$ are the photon annihilation  and creation operators acting on the photon Fock states $|n_{n \mathbf{q} }\rangle $, $\omega_q = c|{\bf q}|$, $V$ is the quantization volume with periodic boundary conditions,  and ${\bf e}_{n{\bf q}}$ is the polarization vector such that ${\bf q}\cdot{\bf e}_{n{\bf q}} =0.$
The total Hamiltonian is 
\begin{align}
\hat{H} = \hat{H}_e + \hat{H}_{\rm ph} + \hat{H}_{int}~,
\end{align}
where the interaction Hamiltonian $\hat{H}_{int} = -\frac{1}{c} \hat {\bf j}  \cdot \hat {\bf A} $ and the current operator is $\hat {\bf j}  = - e v_F{\boldsymbol\sigma}$. 

Now consider the normal incidence of the radiation on the Hall sample, when ${\bf q}= q {\bf z}_0$, and take the field quantization volume as a ray bundle of volume $V = l_xl_yl_z$. For our purpose it is sufficient to consider only one spatial mode at frequency $\omega_q = cq$ and polarization ${\bf e}_q$. It is straightforward to generalize it to a multimode wave packet. For the description of propagating multimode single-photon wave packets, see \cite{PhysRevLett.131.233802}. 

In the rotating-wave approximation (RWA) the interaction Hamiltonian becomes
\begin{align}
\hat{H}_{int} = -\sum_{k,n > n^{\prime}}\sqrt{\frac{2\pi\hbar}{V\omega_{q}}}
\left({\bf e}_{q}\cdot {\bf j}_{nk;n'k}{\hat b}_{q}{\hat a}^{\dagger}_{kn}{\hat a}_{kn^{\prime}} + H.c.\right) ~.\nonumber 
\end{align}
In particular, in the weak-field linear regime of unperturbed populations we obtain  the absorption probability per unit time and per given transition $n'\to n$ as  
\begin{eqnarray} 
{\alpha}^{n'\to n}_{{\bf e}_{q}} = \sum_k\frac{4\pi|{\bf j}_{nk;n'k}\cdot{\bf e}^{*}_{q}|^2}{\hbar\omega(l_xl_yl_z)}\frac{\gamma\left(F^{(0)}_{n^{\prime}k}-F^{(0)}_{nk}\right)}{\gamma^2 + (\omega_{nk;n'k} -\omega)^2}
\label{absorp_dim}
\end{eqnarray}
where $\gamma$ is the homogeneous broadening of inter-LL transitions determined by disorder-induced scattering and $F^{(0)}_{jk} = \left(1+e^{(E_{jk}-E_F)/(k_BT)}\right)^{-1}$ is the Fermi distribution function at temperature $T$. For small $\gamma$ as compared to the inhomogeneous broadening of transitions between edge states resulting from electron dispersion, the Lorentzian in Eq.~\eqref{absorp_dim} becomes the delta function $\pi \delta(\omega_{nk;n'k} -\omega)$ and one recovers the Fermi's golden rule expression. In the case of a finite $\gamma$,  stochastic noise terms appear in the equations for complex amplitudes according to the stochastic Schr\"{o}dinger equation approach \cite{PhysRevA.107.013721}, but this will not affect the linear absorption probability.

The summation over $k$ can be replaced by integration, $\sum_k \to g_s \frac{L_x}{2\pi} \int dk$,   
where  $L_x$ is the quantization length of electron states in x-direction along the edge.  The total absorption probability is obtained by adding the contributions from all LL transitions,  $ {\alpha}^{tot}_{{\bf e}_{q}} = \frac{l_z}{c} \sum_{n,n^{\prime}}{\alpha}^{n'\to n}_{{\bf e}_{q}}$. Here we also converted the probability per unit time into the total dimensionless probability of absorbing a photon by multiplying the former by the photon pulse duration $\Delta t \approx l_z/c$. Defined this way, the photon absorption probability will also describe the dimensionless absorbance of the classical monochromatic wave by a 2D system \cite{singh}. 

One can make the expression in Eq.~\eqref{absorp_dim} independent on the sample and beam geometry by dividing $\alpha^{tot}_{{\bf e}_{q}}$ by the geometric factor  $f = \ell_c L_x/(l_x l_y)$, 
\begin{equation}
{\mathscr A}_{{\bf e}_{q}}(\omega) = \frac{{\alpha}^{tot}_{{\bf e}_{q}}}{f} = \frac{1}{f}  \frac{l_z}{c} \sum_{n,n^{\prime}}{\alpha}^{n^{\prime}\to n}_{{\bf e}_{q}}.
    \label{absor} 
\end{equation}
Clearly, $f$ measures the degree of the overlap between the edge area and the beam area. If the incident radiation is focused on a sample with a nanotip or nanoantenna, one can obtain $f \sim 1$ and significant absorption probability as one can see below.  

In Fig. \ref{absorbance_KKp} we show the geometry independent absorbance spectra for graphene with zigzag edge for four different positions of the fermi level. The regions of electron momenta in the first column contributing to different absorbance peaks are shaded with colors that match the colors of the peaks. The absorbance peaks are well separated. Clearly, by tuning the frequency one can probe a transition between particular edge states or excite electrons into a particular edge state.  

Perhaps the most remarkable fact is that the absorbance peaks corresponding to different valleys are also well separated, especially with increasing Fermi level. Therefore, one can do valley-selective spectroscopy or excite electrons into an edge state with a given valley index. This opens the possibility of manipulating and using valley index degree of freedom in transport, interferometry, and optical experiments and devices. Note also that for the geometric overlap factor $f \sim 1$ the unnormalized absorption probability becomes a sizable fraction of 1 for each peak. When it approaches 1, the linear perturbative treatment breaks down and one has to include the dynamics of edge state populations, which will be considered elsewhere. 

The lowest-energy peak in the polarized absorbance spectra shown in the second and third columns of Fig. \ref{absorbance_KKp}  is dominated by transitions between the bulk states across the Fermi level because of their dispersionless nature. Even though the peaks due to edge states are at significantly higher energies and are well isolated, bulk absorption may still be non-negligible due to disorder-induced broadening. Fortunately, one can get rid of the bulk states absorption completely by subtracting $x$-polarized absorbance from that of $y$-polarized radiation. This will leave us with purely edge contribution.

\begin{figure}[t!]
\includegraphics[width =\linewidth]{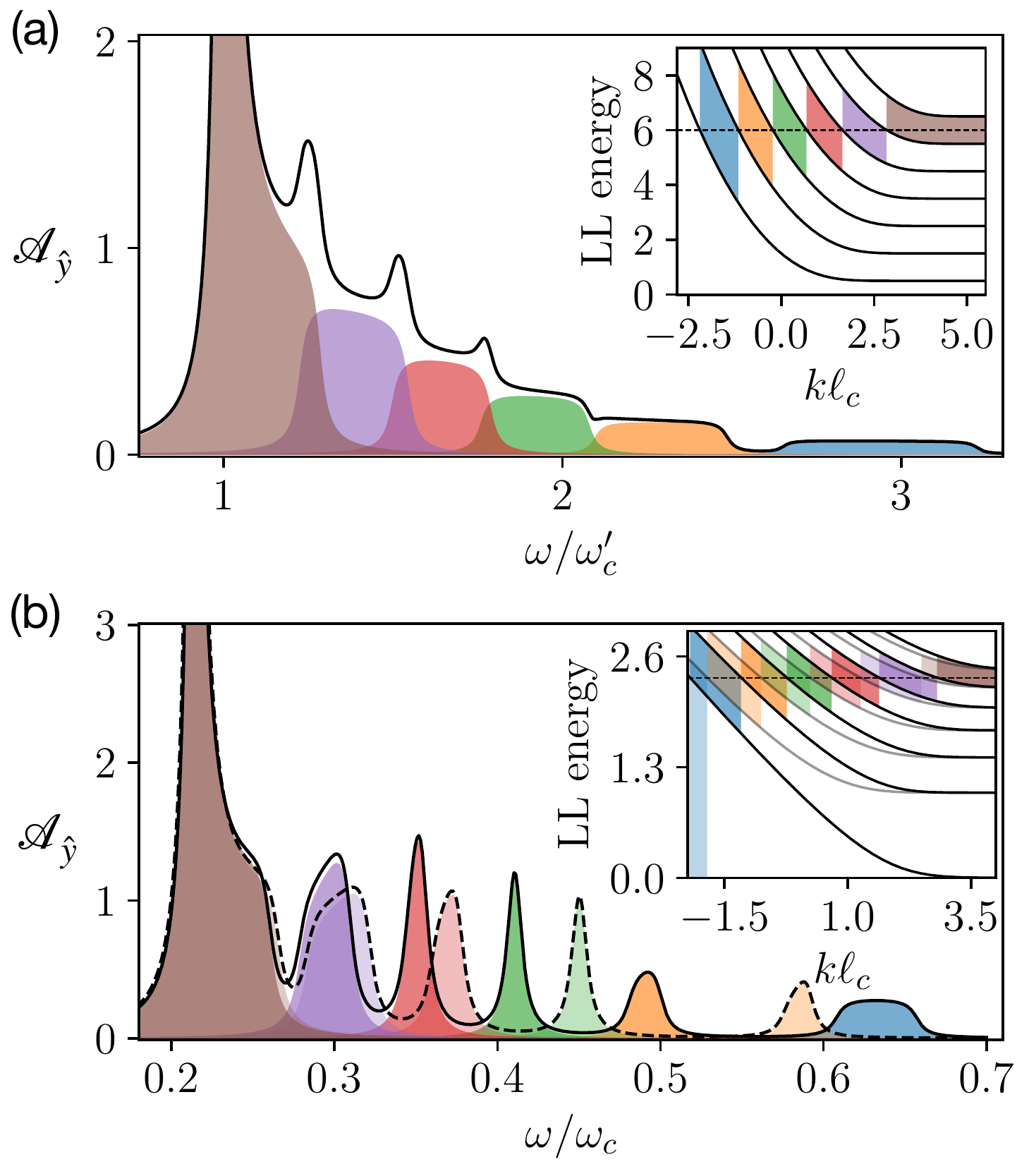}
\caption{Comparison of $y$-polarized absorbance spectra of QH samples for (a) nonrelativistic 2DEG in semiconductor QWs and (b) graphene (zigzag edge). The bulk contribution comes from $n = 5$ to $n = 6$ transition in both cases, for which we have placed the Fermi level between these LLs: (a) $E_F = 6\hbar\omega^{\prime}_c$ and (b) $E_F = 2.35\hbar\omega_c$ where $\omega^{\prime}_c$ and $\omega_c$ as the cyclotron frequencies for these two systems. Colors indicating the transitions between different LLs match the shading of corresponding regions in absorbance spectra. Solid and dashed curves indicate the absorbance due to K and K'-valley states, respectively.}
\label{fig_comp}
\end{figure}

It is instructive to compare the absorbance spectra for graphene with those for nonrelativistic 2DEG in semiconductor QWs. As shown in Fig. \ref{fig_comp}, the contributions from transitions between different LLs to different spectral regions in the absorbance spectra overlap much stronger for the nonrelativistic 2DEG as compared to graphene where the peaks corresponding to different edge states and different valleys are essentially isolated. Therefore, graphene seems to be an ideal platform to control the LL index and valley index of edge state excitations.

\section{Optical rectification and edge photocurrent}\label{DC_gen}
Without light, moving electrons in chiral edge channels constitute a dissipationless DC current, which for a fixed valley in QH samples can be only due to the diagonal elements of the density matrix, 
\begin{align}
\label{J_diag}
\mathcal{J}_0 = -\frac{e}{L_x} \sum_{n,k}\left(v_x\right)_{nk;nk}\rho^{(0)}_{nk;nk}~, 
\end{align}
where $\left(v_x\right)_{nk;nk} = v_F\la \Psi_{nk}|\sigma_x|\Psi_{nk}\ra$ and $\rho^{(0)}_{nk;nk} = F^{(0)}_{nk}$. 
The net current from both edges becomes nonzero if a DC voltage $\Delta V$ is applied. 

Inversion symmetry breaking for edge states enables strong second-order optical nonlinearity which becomes allowed in the electric dipole approximation. This gives rise to the possibility of generating the DC or quasi-DC current  by an {\it optical} field through the second-order process of the optical rectification, mediated by corresponding off-diagonal elements of the density matrix.  In order to support the net current only one edge needs to be illuminated; otherwise the contributions of two opposite edges will still cancel each other. 

The derivation of the optically driven DC current is the same as for the nonrelativistic 2DEG \cite{singh}, although the current magnitude and spectral distribution are very different. Here we outline the main steps for completeness.  The general expression for the DC current is 
\begin{align}
\mathcal{J}_{\rm dc} = -\frac{e}{L_x} \sum_{n\neq m,k} \left(v_x\right)_{nk;mk}\rho_{mk;nk}~,
\label{rect1} 
\end{align}
where $\left(v_x\right)_{nk;mk} = v_F\la \Psi_{nk}|\sigma_x|\Psi_{mk}\ra$. Here we again limit ourselves to the transitions inside the conduction band. 
The density matrix elements are found by solving the master equation in the perturbative series with respect to the interaction Hamiltonian $\hat V(t) = ey({\mathcal E}e^{-i\omega t}+{\mathcal E}^*e^{i\omega t})$, where we assume the field to be classical in this section:
\begin{align}
\frac{\partial \rho^{(j+1)}_{nm}}{\partial t} =  - i\omega_{nm}{\rho}^{(j+1)}_{nm} -\frac{i}{\hbar}\sum_{\nu}\left(V_{n\nu}\rho^{(j)}_{\nu m} - \rho^{(j)}_{n\nu}V_{\nu  m}\right),
\end{align}
with $n (m)$ denoting the quantum state with a given momentum corresponding to Landau level index $n (m)$ for a given $s(s^{\prime})$. Here we suppressed the electron momentum in the subscripts, although the final result for the current has to be integrated over electron momenta, as in Eq.~\eqref{rect1}. Interactions and impurity effects will be considered at a phenomenological level for simplicity by adding a damping term $\gamma \rho_{nm}$. To the first order in ${\mathcal E}$,  
\begin{align}\label{coherence_alpha_beta}
{\rho}^{(1)}_{nm} =  -\frac{e}{\hbar}\frac{y_{nm}\left(\rho^{(0)}_{mm} - \rho^{(0)}_{nn}\right)}{\omega_{nm} - \omega - i\gamma}~\mathcal{E}e^{-i\omega t}~,
\end{align}
where ${\rho}^{(1)}_{mn} = \left({\rho}^{(1)}_{nm}\right)^{*}$ and $y_{nm}=\langle \Psi_{nk, s}|y|\Psi_{mk, s^{\prime}}\rangle$ and we have suppressed the valley index.

The second-order nonlinear density matrix elements contributing to the rectification current satisfy the equations of motion
  \begin{eqnarray}\nonumber
		&&\frac{\partial \rho^{(2)}_{n;(n-1)}}{\partial t} + i\omega_{n;(n-1)}\rho^{(2)}_{n;(n-1)} = -\frac{ie }{\hbar} {\mathscr F}(t) \times\\\nonumber
		&&\left[ y_{n;(n-2)}\rho^{(1)}_{(n-2);(n-1)} - y_{(n-2);(n-1)}\rho^{(1)}_{n;(n-2)}\right.\\\nonumber
		&&+y_{n;(n+1)}\rho^{(1)}_{(n+1);(n-1)} - y_{(n+1);(n-1)}\rho^{(1)}_{n;(n+1)}\\
		&&\left.+(y_{n;n} - y_{(n-1);(n-1)})\rho^{(1)}_{n;(n-1)}\right],
\end{eqnarray}
  \begin{eqnarray}\nonumber
		&&\frac{\partial \rho^{(2)}_{(n+1);(n-1)}}{\partial t} + i\omega_{(n+1);(n-1)}\rho^{(2)}_{(n+1);(n-1)} = \\\nonumber
		&&-\frac{ie}{\hbar}{\mathscr F}(t) \left[ y_{(n+1);n}\rho^{(1)}_{n;(n-1)} - y_{n;(n-1)}\rho^{(1)}_{(n+1);n}\right.\\
		&&\left.+(y_{(n+1);(n+1)} - y_{(n-1);(n-1)})\rho^{(1)}_{(n+1);(n-1)}\right],
\end{eqnarray}
where we have defined ${\mathscr F}(t) = {\mathcal E}e^{-i\omega t}+{\mathcal E}^*e^{i\omega t}$.
Time-independent solution to these equations under illumination with a monochromatic field gives rise to a large number of terms contributing to the rectification current, which grows rapidly with increasing doping. However, all terms either scale as  $y_{n;(n+2)}\rho^{(2)}_{(n+2);n}$ or as $y_{n;(n+1)}\rho^{(2)}_{(n+1);n}$ where $ \rho^{(2)}_{(n+1);n} \propto (y_{(n+1);(n+1)} - y_{n;n})$.  As one can see from Fig.~\ref{dipole}, both $y_{n;(n+2)}$ and $(y_{(n+1);(n+1)} - y_{n;n})$ are only nonzero for electron states close to the edge, when $y_k$ is within several magnetic lengths $\ell_c$ from the edge.

\begin{figure}[t!]
\includegraphics[width =\linewidth]{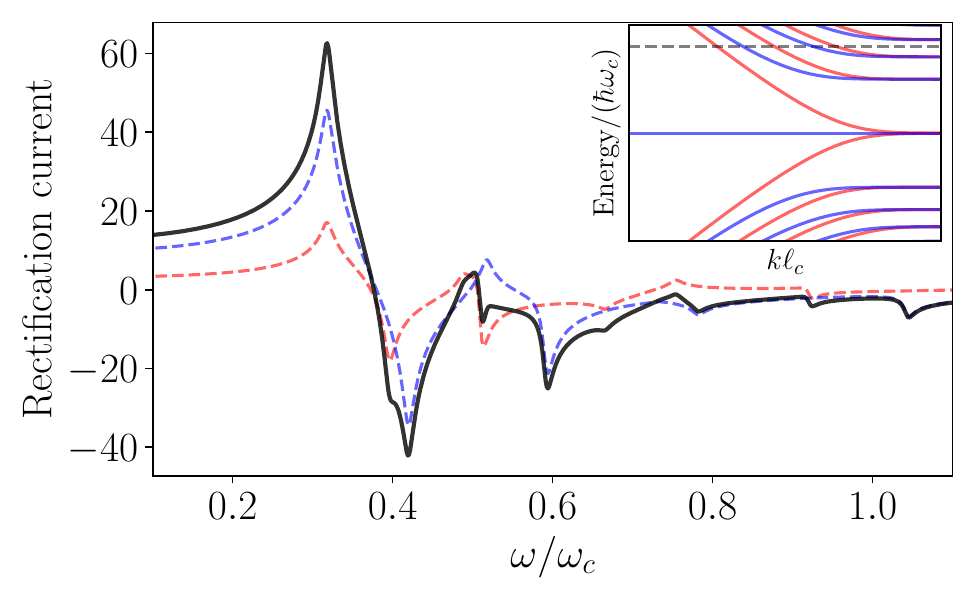}
\caption{Normalized rectification current from Eq.~\eqref{normcur}  as a function of the normalized optical frequency  for $\Gamma = 0.005\, \omega_c$. Contribution to the current from $K$ and $K^{\prime}$ valleys are shown in dashed red and dashed blue lines respectively. Solid black line is the sum of these contributions. The Fermi level $E_F = 1.6\, \hbar\omega_c$ lies between $n = 2$ and $n = 3$ bulk LLs  (inset).}
\label{non_linear}
\end{figure}

It is convenient to normalize the rectification current as
\begin{align}
\tilde J_{i} = \left(\frac{\mathcal{J}_{\rm dc}}{\zeta\mathcal{J}_{0}}\right)_{i}~,
\label{normcur}
\end{align}
where we keep only the second-order contributions in Eq.~\eqref{rect1}, $i$ is the valley index, and 
\begin{align}
\zeta = \frac{e^2|{\mathcal E}|^2\ell_c^2}{\hbar^2\omega_c^2}
\end{align}
is the nonlinearity parameter which has the same order of magnitude as the ratio of the square of the Rabi frequency of the optical field to the cyclotron frequency squared \cite{singh}. As an example, for a magnetic field strength of $1$T, the nonlinearity parameter becomes equal to 1 for an electric field amplitude $|{\mathcal E}| \approx  1.4\times 10^6$ V/m. In this case the optical pumping to excited LLs becomes significant. Obviously we need to keep the intensity lower so that $\zeta \ll 1$ and the perturbative calculations are justified. Still, the rectification current can exceed the equilibrium chiral current in certain frequency ranges. 

In Fig.~\ref{non_linear}, we show the valley resolved as well as total rectification current spectra for the Fermi level $E_F = 1.6\, \hbar\omega_c$ which falls within the gap between the bulk $n = 2$ and $n = 3$ LLs as shown in the electron energy vs. momentum plot in the inset. The strongest peak near $\omega\approx 0.3\, \omega_c$ stems from the non-vanishing $(y_{33}-y_{22})y_{32}^2/(\omega_{32}-\omega-i\gamma)$ term near the edge. It changes sign above resonance $\omega > \omega_{32}$ as expected. At even higher frequencies, the contributions from other LL transitions become noticeable. With increasing doping level the main peak position is shifted to lower frequencies, and vice versa, which is a consequence of nonequidistance of LLs in graphene. Furthermore, the magnitude of the current is much higher than in the case of nonrelativistic electron dispersion \cite{singh}. 
 
%

\section{Armchair edge}
\label{armchair}

\begin{figure}[t!]
\includegraphics[width =\linewidth]{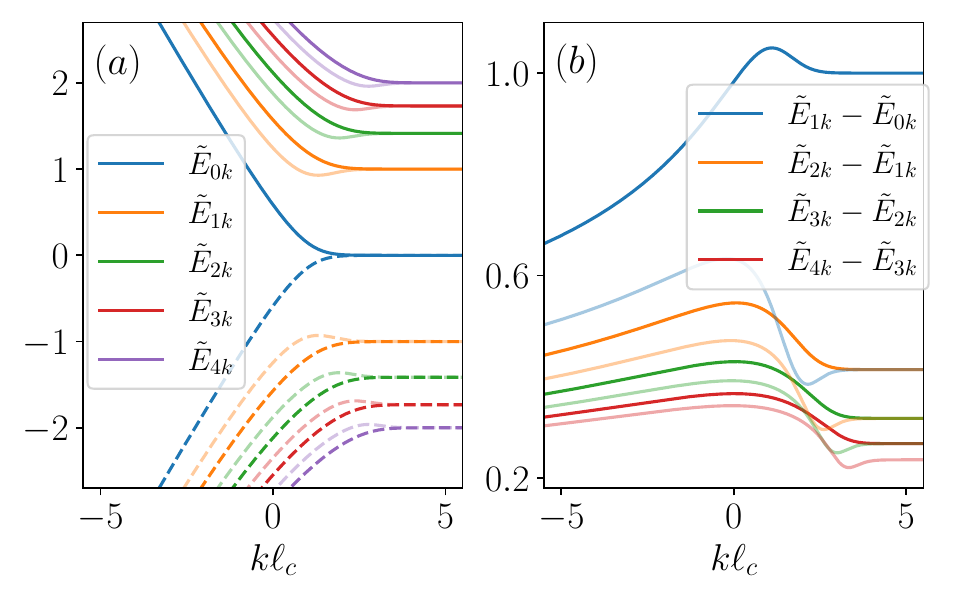}
\caption{(a) Electron energies and (b) transition energies between adjacent states (normalized with $\hbar\omega_c$) near the armchair edge  as a function of momentum.}
\label{dispersion_armchair}
\end{figure}

In this case, the overall qualitative behavior of the absorbance spectra and the nonlinear response is similar to the zigzag edge, but quantitative details are quite different. The main distinction comes from the fact that for an armchair edge both sublattices meet the boundary simultaneously. Therefore, a linear combination of wavefunctions from $K$ and $K^{\prime}$ valleys must vanish at the edge \cite{PhysRevB.73.195408}. In this case it is simpler to work in the $A=Bx \hat y$ gauge and take the edge at $x = 0$. As compared to the zigzag edge, mixing of valley wavefunctions leads to characteristically 
distinct LL energies for the armchair edge states, as shown in Fig. \ref{dispersion_armchair}. Unlike the zigzag case, where the degeneracy between $K$ and $K^{\prime}$ valley is lifted for the edge states, here LL energies bifurcate due to valley mixing. This applies to all $E_{nk}$, except $n=0$ as shown in Fig.~\ref{dispersion_armchair}(a), where the light colors are used for the lower energy branches.  Note also that the dispersionless zero energy state disappears. Furthermore, odd energy states first lower their energies as compared to the bulk states before increasing monotonically as electrons move closer and closer to the edge. This leads to a much more pronounced non-monotonic dispersion of the transition energies in Fig.~\ref{dispersion_armchair}(b) as compared to Fig.~\ref{dispersion}(b). 
%
%
%
%
In the case of an armchair edge, 
%
%
%
%
%

For the Fermi level $E_F>0$, we will again focus on the transitions in the conduction band in order to calculate the absorbance spectra. Their wavefunctions can be written as \cite{Gusynin2008, PhysRevResearch.3.013201}
\begin{align}
    \Psi^{AC}_{nk}(x,y) = \frac{e^{iky}}{\sqrt{L_y C^{\prime}_{n}}}
    \begin{pmatrix}
        -i\tau_n D_{n}(x^{\prime})e^{iKx}\\
        -\tau_n\sqrt{n}D_{n-1}(x^{\prime})e^{iKx}\\
        i\sqrt{n}D_{n-1}(x^{\prime})e^{-iKx}\\
        D_{n}(x^{\prime})e^{-iKx}
    \end{pmatrix}~,
\end{align}
where $x^{\prime} = \sqrt{2}(x/\ell_c + k\ell_c)$, $k$ is now the $y$-component of the electron wave vector counted from $K$ and $K'$  valley positions at wavevectors $K$ and $-K$, $L_y$ is the quantization length for electron states along $y$-direction, $C_n^{\prime}$ is the normalization constant for the spinor and $\tau_n = (-1)^{n+1}$. The two top (bottom) components are contributions from the $K$ ($K^{\prime}$) valley.
%
%
The current operator is 
\begin{align}
    \hat {\bf j} = ev_F\left[\begin{pmatrix}
        \sigma_x &0\\
        0 & -\sigma_x
    \end{pmatrix}, \begin{pmatrix}
        \sigma_y &0\\
        0 & \sigma_y
    \end{pmatrix}\right].
\end{align}
The $x$ and $y$-components of the current matrix elements for transitions involving $n$ and $m$-th LLs are given by 
\begin{eqnarray}
    \langle \chi_{n}|\hat j_x|\chi_{m}\rangle &=& (1+\tau_n\tau_m)\frac{iev_F}{\sqrt{C^{\prime}_{n}C^{\prime}_{m}}} ~\mathcal{I}^{-}~,\\
    \langle \chi_{n}|\hat j_y|\chi_{m}\rangle &=& -(1+\tau_n\tau_m)\frac{ev_F}{\sqrt{C^{\prime}_{n}C^{\prime}_{m}}} ~\mathcal{I}^{+}~,
\end{eqnarray}
where
\begin{align}
\mathcal{I}^{\mp} =\int_{-\tilde x_{\rm max}}^0d\tilde x\left[\sqrt{n}D_{n-1}D_{m} \mp \sqrt{m}D_{m-1}D_{n}\right],
\end{align}
where $\tilde x = x/\ell_c$, $\tilde{x}_{\rm max}$ plays the role of $\tilde{y}_{\rm max}$ in Eq.~\eqref{dipole_mat} and is taken to be $20$. It is interesting to note that while in the zigzag case the transitions between $K$ and $K^{\prime}$ valleys are absent,  in the armchair case due to valley mixing (hybridization) the selection rule is determined by the factors $\tau_{n,m}$, namely only those transitions are allowed for which $\tau_{n}\tau_{m}\neq -1$.


\begin{figure}[t!]
\includegraphics[width =\linewidth]{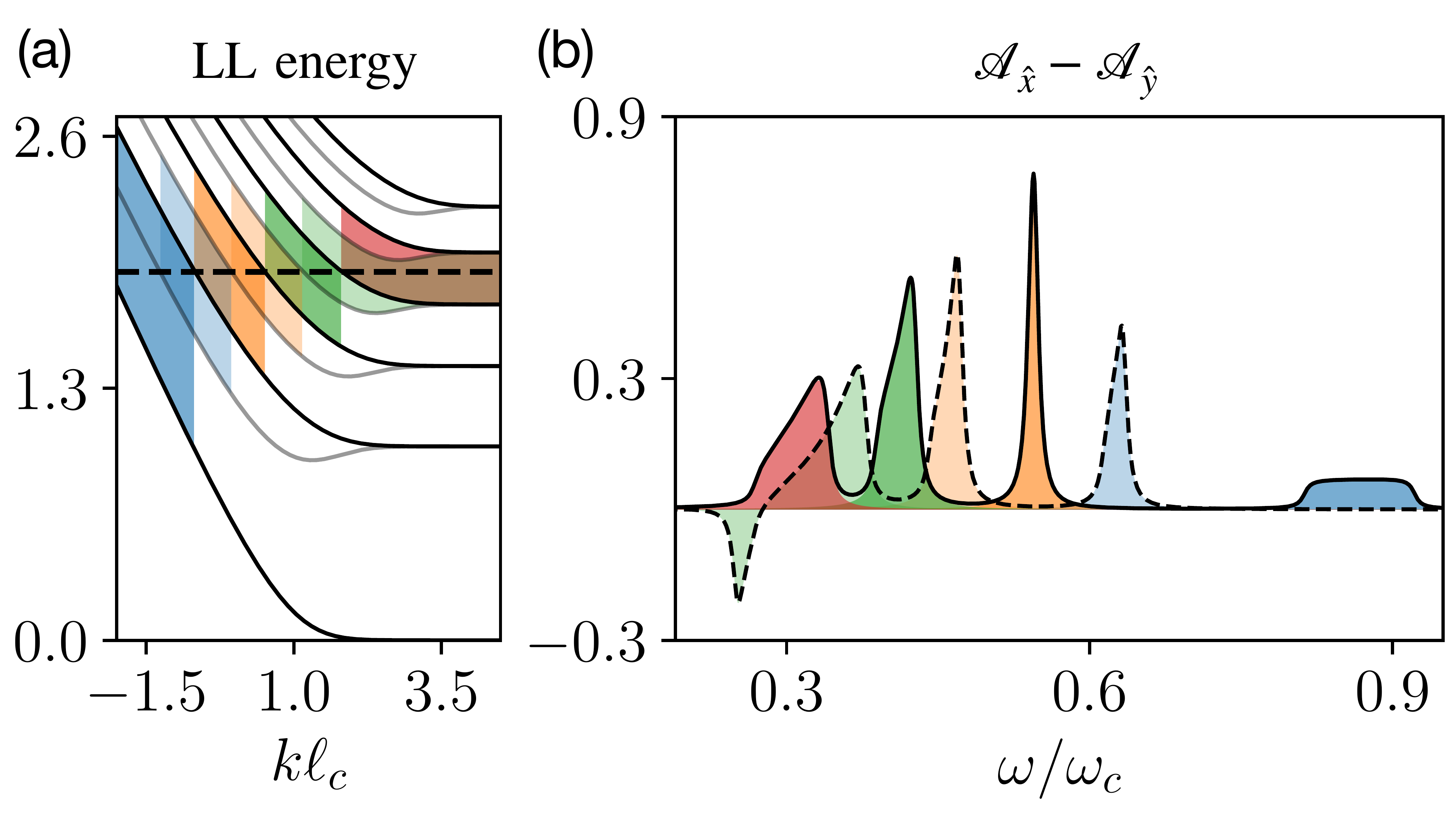}
\caption{(a) Electron energies normalized by $\hbar \omega_c$ near the armchair edge with the Fermi level at $E_F = 1.9\, \hbar\omega_c$ shown with dashed line. (b) The difference between normalized 2D absorbances of $x$ and $y$-polarized radiation. Regions in momentum space in (a) contributing to different absorbance peaks in (b) are shaded with colors that match the colors of the peaks.  $\Gamma = 0.005\, \omega_c$.}
\label{absorption_armchair}
\end{figure}

Proceeding with photon absorption probability calculations in the same way as in the case of a zigzag edge, we obtain geometry-independent normalized absorbances for $x$ and $y$ polarized radiation defined as in Sec.~III.  
In Fig.~\ref{absorption_armchair}(b) we show the difference between the normalized absorbances for $x$ and $y$ polarized light for the armchair case.  Similarly to the zigzag edge, the absorbance spectra consist of well-separated peaks in frequency domain which can be uniquely mapped to the transitions between particular LL edge states as can be seen from Fig.~\ref{absorption_armchair}(a). Note also the sign reversal of the absorbance difference for at least one transition. This effect is in general doping dependent.  

The calculations of the rectification current in the armchair case result in qualitatively similar current magnitudes and spectra, and we do not show them here. 


%

\section{Conclusions}

In conclusion, 
we have demonstrated the feasibility of optical spectroscopy and valley-selective coherent optical control of chiral edge state populations and currents under the conditions of the integer QH effect in graphene. 

The excitations of edge states with a given LL index are possible with single-electron sensitivity, especially when  the  spatial overlap of incident photon modes with the edge state area is enhanced through nanofocusing with a tip or nanoantenna.  Since the optical transitions between edge states have significantly different transition energies and polarization selection rules as compared to the bulk of the sample,  selective excitation of a specific edge channel is possible without disturbing the rest of the sample. Moreover, valley-selective optical control of edge states is possible for a zigzag edge, due to the fact that absorbance peaks for different valleys are well separated in frequency. 

Furthermore, inversion symmetry breaking near the sample boundary enables strong second-order optical nonlinearity in the electric dipole approximation, resulting in efficient optical rectification of incident radiation and direct optical driving of a quasi-DC current in edge states. 

The predicted optical effects can be used to study or control edge currents by optical means, to control the interference pattern in QH interference experiments, and maybe even to manipulate LL index and valley index as degrees of freedom in electron qubits in  QH interferometers. We hope that this study will foster collaboration between the QH and optical communities.

\section{Acknowledgements} 
This work has been supported in part by the 
National Science Foundation Award No. 1936276 and the Air Force Office for Scientific Research Grant No. FA9550-21-1-0272. M. T. acknowledges the support by the Center for Integration in Science of the Ministry of Aliya and Integration, Israel.

\bibliography{Ref2}
\end{document}